\documentclass[applsci,article,accept,moreauthors,pdftex]{mdpi} 
\usepackage{color} 
\usepackage{graphicx}
\usepackage{caption}
\usepackage{sidecap}
\usepackage{float} 
\usepackage{amssymb}
\usepackage{amsfonts}
\usepackage{hyperref}
\usepackage[hypcap=true]{caption}
\usepackage[labelformat=simple]{subcaption}

\firstpage{1} 
\pubvolume{xx}
\issuenum{1}
\articlenumber{5}
\pubyear{2019}
\copyrightyear{2019}
\history{Received: November 29, 2019; Accepted: March 02, 2020 Published:
  March 06, 2020}

\Title{Quantum Many-Body Theory for Exciton-Polaritons in
  Semiconductor Mie Resonators in the Non-Equilibrium}

\Author{Andreas Lubatsch $^{1\dagger}$ and Regine Frank $^{2,3\dagger}$*}

\AuthorNames{Andreas Lubatsch  and Regine Frank}

\address{%
$^{1}$ \quad Georg-Simon-Ohm University of Applied Sciences, Ke{\ss}lerplatz
  12, 90489 N\"urnberg, Germany\\
$^{2}$ \quad Bell Labs,\,600 Mountain Ave., Murray Hill,\,New Jersey\,07974\\
$^{3}$ \quad Serin Physics Laboratory, Department of Physics and Astronomy,
Rutgers University, 136 Frelinghuysen Road, Piscataway, NJ 08854-8019, USA; regine.frank@rutgers.edu}

\corres{Correspondence: regine.frank@googlemail.com}

\firstnote{These authors contributed equally to this work.}

\abstract{We implement externally excited ZnO Mie resonators in a framework of a generalized Hubbard Hamiltonian to investigate
the lifetimes of excitons and exciton-polaritons out of thermodynamical
equilibrium.  Our results are derived by a Floquet-Keldysh-Green's formalism with Dynamical Mean Field
Theory (DMFT) and a second order iterative perturbation theory solver (IPT). We find that the Fano resonance which originates from coupling
of the continuum of electronic density of states to the semiconductor Mie
resonator yields polaritons with lifetimes between $0.6$ ps and $1.45$
ps. These results are compared to ZnO polariton lasers and to ZnO random
lasers. We interpret the peaks of the exciton-polariton lifetimes in our
results as a sign of
gain narrowing which may lead to stable polariton lasing modes in the single
excited ZnO Mie resonator. This form of gain may lead to polariton random lasing in an
ensemble of ZnO Mie resonators in the non-equilibrium.}



\begin{document}

\setcounter{section}{-1} 

\section{Introduction}
\label{INTRO}

\begin{SCfigure}
\centering
 \vspace*{-0.8cm}\includegraphics[width=8.0cm]{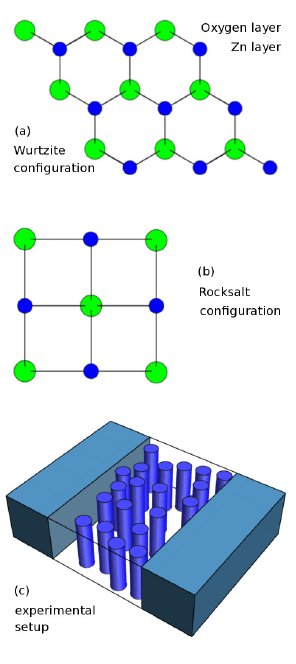}
\caption{ZnO structure (ab-plane). {\bf (a)} Non-centrosymmetric, hexagonal, wurtzite
    configuration {\bf (b)}
    Centrosymmetric, cubic, rocksalt configuration (Rochelle salt)
    \cite{JMATCHEMA,Rocksalt1,Rocksalt2}. The rocksalt configuration is
    distinguished by a tunable gap from $1.8\,eV$ up to $6.1\,eV$, a gap value
    of $3.38\,eV$ is typical for the  monocrystal rocksalt configuration without
    oxygen vacancies \cite{ZnOMono1,ZnOMono2}. As such the rocksalt
    configuration could be suited for higher harmonics generation
    under non-equilibrium topological excitation \cite{LubatschEPJ2019,Symmetry,Faisal}. {\bf (c)} Typical setup of
    a solid state random laser: Densely packed disordered ZnO nano-resonators
    are embedded in a waveguide. The sample is externally pumped in (111) direction. Mie
    resonances \cite{FrankPhysRevResearch} occur at the inner cylinder jacket and they couple to the
    density of bulk states in the sense of a Fano resonance. We consider in this article the exciton and exciton
polariton dynamics of the single scatterer as the building block of the solid
state random laser or the polariton random laser. Laser gain is attributed to
the accumulation of excited states at a certain level along  with a
significantly increased life-time of the excited state, i.e. a stable laser in
stationary state. In the random laser each scatterer of a
well defined area or volume as part of an ensemble of several scatterers is
going to lase. The investigation of the complex Mie-resonance in the
non-equilibrium as the origin of pronounced laser levels for polariton lasing
is on target. The formation of the polariton as a quasiparticle is
characterized by the coupling strength $\gamma\,=\,g/t$ of the bulk electronic procedures and the cavity resonance which
is on the order of the band width or one order below and thus it is in the polariton range
\cite{Lamata}. The
waveguide principally plays only a subordinate role, it does not
serve as the dominating conventional laser resonator cavity.}
\label{ZnOLAttice}
\end{SCfigure}

Semiconductor micro-cavities operating in the strong coupling regime are
optical resonators in which the {\em eigenmodes} are no longer purely
excitonic or photonic, but a mixed light matter state is occurring. This
evolution has finally led to polariton lasing
\cite{YaNature,Kaliteevski,YaScience,Imamoglu,Dang,Hoefling,Li,Guillet,Bajoni}. Random
lasing \cite{Cao} has been claimed to be fundamentally incompatible with
polaritons. From recent research developments however it turns out that it is
not yet clear whether random lasers are either pure photon lasers or pure
polariton lasers, or whether they even have both characteristics at the same
time \cite{Sasaki}. Thresholds like in conventional lasers may be observed \cite{Deng1,Deng2}.
We show in this article that for excitations of the quantum many-body system
of ZnO nano-resonators by experimentally feasible threshold intensities of
the optical pump
found in  
random laser experiments we derive theoretically well confined peaks in terms of gain
narrowing in the spectrum of the
lifetime for the cavity polariton in the non-equilibrium. These peaks give evidence
that such a sample may undergo a transition towards exciton-polariton lasing rather
than a transition to exciton-photon lasing, at least a coexistence of both regimes may occur.
ZnO is known for excellent light-matter coupling characteristics
\cite{LubatschEPJ2019,Symmetry,Lu,Petrosyan,Lu2,Shimada,Li,SUN,DUAN,JMATCHEMA,ZnOMono1,ZnOMono2,Rocksalt1,Rocksalt2,vanBlaaderen}, it is predestined for this
study.  It has been experimentally derived that ZnO can be
  considered as a Mott insulator under certain conditions
  \cite{Lu,Lu2}. A feature of ZnO nano-structures is a transition from the
  non-centrosymmetric wurtzite structure to the centrosymmetric rocksalt
  configuration with a variation of temperature and pressure, see Fig. \ref{ZnOLAttice}. When the
system is externally pumped, Fig. \ref{ZnOLAttice}(c), lasing may occur eventually. The propagating light intensity
in the random laser may experience Mie resonances as a whispering gallery
resonance of light at the inner surface of the individual nano-structure. 
In a
complex medium, which pumped ZnO in the non-equilibrium certainly is, the
resonant Mie mode forms a light matter bound state in the form of a Fano
resonance. It leads to quantum many body physics and the renormalization of
the bands under excitations. We consider {\bf (i)} the semiconductor material as it is
subjected to the strong external AC field of the optical pump in
  the sense of a topological excitation \cite{LubatschEPJ2019,Symmetry}. We
  model the coupling of the classical laser, $\hbar\Omega_L\,=\,1.75\,eV$, to the quantum
many body system as well as higher-order photon absorption processes,
$\hbar\Omega_L\,=\,3.5\,eV$ etc., by means of the Floquet matrix
\cite{Floquet,Grifoni,PRB,FrankANN}. The band structure and the lifetimes of 
light-matter coupled states are derived by the Dynamical Mean Field Theory
(DMFT)
\cite{Kotliar,ANN,NJP,APLB}
in the sense of a generalized Hubbard-Hamiltonian out-of thermodynamical equilibrium. The Hubbard Hamiltonion includes strictly local
  interactions between electrons with opposite spins. It is predestined
  to describe the exciton dynamics of excited matter in a physical time range
  where additional impurity scattering is not expected. Bulk ZnO develops under excitation in the near band
  gap region an exciton dynamics. 
In the second step of this work {\bf (ii)} we additionally consider photons which populate 
the  Mie resonance of the individual nano-resonator and we study the
formation of polaritons. The spectral features are compared to experimental
emission spectra of ZnO random lasers and polariton lasers and
we find a very good agreement in either case.

\section{Theory}
\subsection{Hubbard Model for Excitons and Exciton-Polaritons}
\begin{figure*}[t!]
\centering
{\rotatebox{0}{\includegraphics[width=1.0\linewidth]{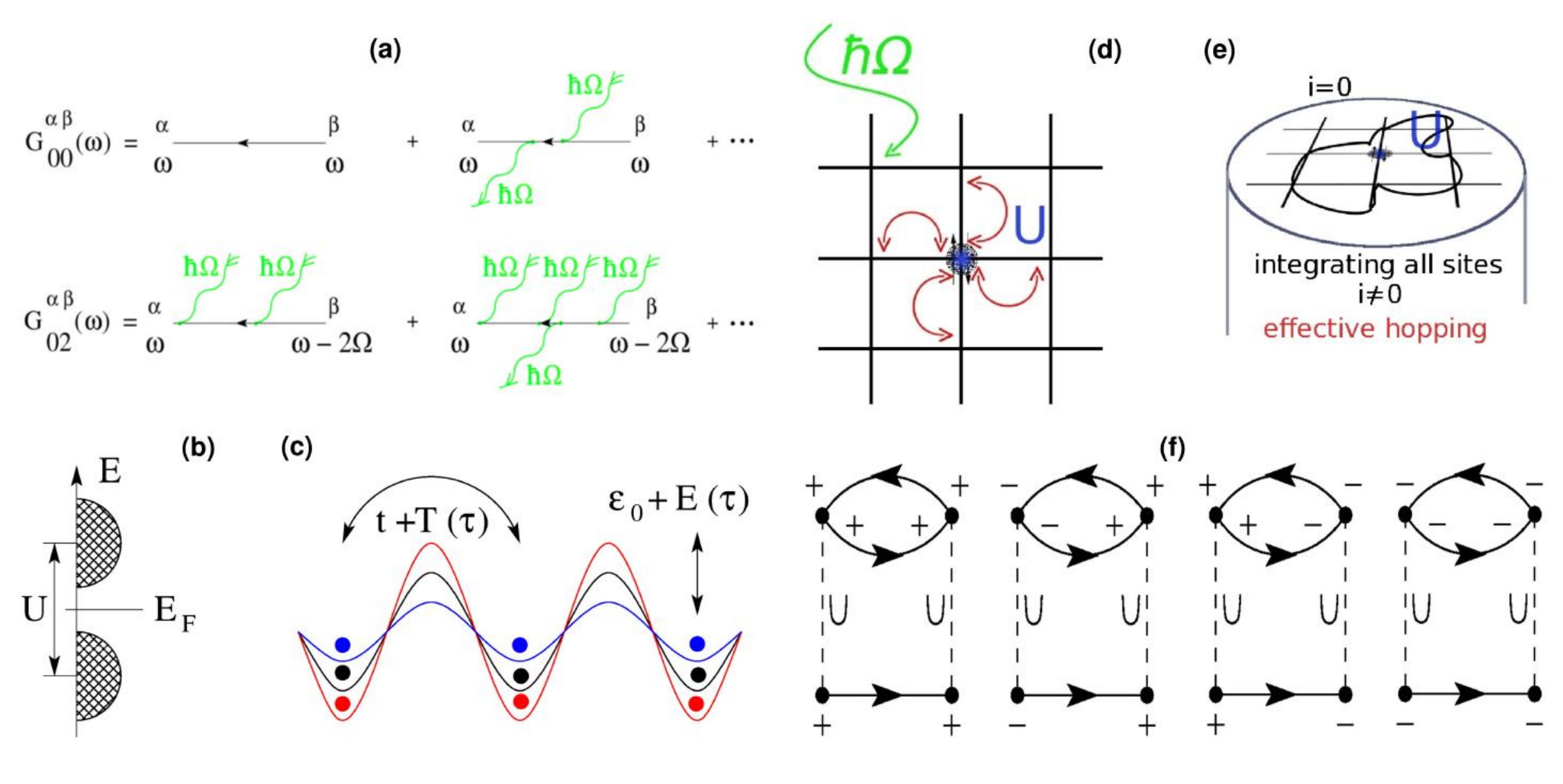}}}
{\caption{Theoretical setup {\bf (a)}  Floquet expansion of the Green's function $G$ into higher
  harmonics of $\hbar \Omega_L$. {\bf (b)} Hubbard model of the electronic system. The gap
    between bands is defined by the
    Coulomb interaction $U$ symmetrically to the Fermi Energy $E_F$. {\bf (c)}
    Tight binding system of the crystal lattice. The hopping parameter
  $t$ is renormalized by external time dependent excitations
  $T(\tau)$. $\varepsilon_0 + E(\tau)$ are renormalizations of the local potential alone. {\bf
    (d)} Non-equilibrium DMFT. Pump photons $\hbar \Omega_L$ enhance electronic
  hopping which is mapped on the single site on the background of the
  surrounding identical lattice sites, the bath. {\bf (e)} Integration over
  all sites
  yields an effective self-consistent theory including non-equilibrium
  effects. The electronic system is coupled to Mie resonances. {\bf
    (f)} Diagrammatic contributions of the second order IPT.}\label{setup}}
\end{figure*}

The interacting Hamiltonian for
the complex driven semiconductor resonator can be written as follows

\begin{eqnarray}
\!\!\!\!\!\!\!\!H\!&=&\!\! \sum_{i, \sigma} \! \varepsilon_i  c^{\dagger}_{i,\sigma}c^{{\color{white}\dagger}}_{i,\sigma} 
+   \frac{U}{2} \sum_{i, \sigma}
c^{\dagger}_{i,\sigma}c_{i,\sigma}c^{\dagger}_{i,-\sigma}c_{i,-\sigma}\label{Hamilton_we}
-  t\!\! \sum_{\langle ij \rangle, \sigma}\!\!
c^{\dagger}_{i,\sigma}c^{{\color{white}\dagger}}_{j,\sigma}\label{Hamilton_we}
\\& +& i\vec{d}\cdot\vec{E}_0 \cos(\Omega_L \tau)\sum_{<ij>,\sigma} 
 \left(
           c^{\dagger}_{i,\sigma}c^{{\color{white}\dagger}}_{j,\sigma} 
 	  -
           c^{\dagger}_{j,\sigma}c^{{\color{white}\dagger}}_{i,\sigma} \right)
 + \hbar\omega_o  a^{\dagger}a^{{\color{white}\dagger}} + g\! \sum_{i, \sigma}
c^{\dagger}_{i,\sigma}c^{{\color{white}\dagger}}_{i,\sigma}  (a^{\dagger}  \!
+ a).\nonumber\\ {\color{white}.}\nonumber
\end{eqnarray}
The electrons are described as a tight binding model, see Fig. \ref{setup},
where the
splitting into valence and conduction band symmetrically to the Fermi level is
included by means of the Coulomb interaction $U$, see
Fig. \ref{setup}(b). The ZnO gap assumes the value of $3.38\,eV$
\cite{Lu,Petrosyan}. The first term in the Hamiltonian $\sum_{i, \sigma}
\! \varepsilon_ic^{\dagger}_{i,\sigma}c^{{\color{white}\dagger}}_{i,\sigma}$  denotes the
local onsite potential. The term $\frac{U}{2} \sum_{i, \sigma}
c^{\dagger}_{i,\sigma}c_{i,\sigma}c^{\dagger}_{i,-\sigma}c_{i,-\sigma}$ is devoted
to the onsite Coulomb interaction $U$ between electrons with opposite
spins. The third term $-t \sum_{\langle ij \rangle, \sigma}\!\!
c^{\dagger}_{i,\sigma}c^{{\color{white}\dagger}}_{j,\sigma}$ is due to the
hopping processes with the amplitude $t$ between nearest neighbor sites. The
classical external pumping is described  in terms of the time dependent field $\vec{E}_0$ with the laser frequency $\Omega_L $, and $\tau$
respectively. In the term $ i\vec{d}\cdot\vec{E}_0 \cos(\Omega_L \tau)\sum_{<ij>,\sigma} 
 \left(
           c^{\dagger}_{i,\sigma}c^{{\color{white}\dagger}}_{j,\sigma} 
 	  -
           c^{\dagger}_{j,\sigma}c^{{\color{white}\dagger}}_{i,\sigma}
         \right)$ is noted the renormalization of the hopping processes due to
         interaction with the pump field. The electronic dipole operator
         $\hat{d}$ is given with strength $|\vec{d}|$. The photonic cavity
         mode $\hbar\omega_oa^{\dagger}a^{{\color{white}\dagger}}$ with the resonance frequency
         $\omega_0$ is coupled to the electron system as $g\! \sum_{i, \sigma}
c^{\dagger}_{i,\sigma}c^{{\color{white}\dagger}}_{i,\sigma}  (a^{\dagger}  \! +
a)$ with a coupling strength $g$ in units of the standard
           hopping $t$. The single band effective Hubbard model
           has been proven to be perfectly suited for the description of the
           movement of hole-bound electrons, excitons,
           as a quasiparticle through the lattice of metal ions
           \cite{Zhang,Jarell,Eder}, which is a lattice of Zn$^{2+}$ ions
           here. For the solution of the Hamiltonian for
         driven bulk matter {\bf (i)} as
         well as for the bulk-cavity coupled system {\bf (ii)} including the
         fifth and the sixth term of Eq. (\ref{Hamilton_we}), the explicit time
dependence of the external field has to be accounted for as well as the
dynamics of the system. It yields Green's
functions which depend on two separate time arguments. The double Fourier
transform from time to frequency coordinates leads to two separate frequencies
which are chosen as relative and center-of-mass frequency
\cite{PRB,FrankANN,NJP,APLB} and we can thus expand the underlying physical
procedures into Floquet modes, the graphical explanation is found in Fig. \ref{setup}(a),
\begin{eqnarray}
\label{Floquet-Fourier}
\!\!\!\!\!\!\!\!\!\!\!\!\!\!\!\!\!\!\!\!\!\!\!\!\!\!\!\!\!\!\!\!\!\!\!\!\!\!\!\!\!\!\!\!\!\!\!\!G_{mn}^{\alpha\beta}(\omega) &=&\!\!\!\left\lmoustache \!\!{\rm d}{\tau_1^\alpha}\!{\rm d}{\tau_2^\beta}\right.
e^{-i\Omega_L(m{\tau_1^\alpha}-n{\tau_2^\beta})}
e^{i\omega({\tau_1^\alpha}-{\tau_2^\beta})}
G (\tau_1^\alpha,\tau_2^\beta)\\
\!\!\!&\equiv& G^{\alpha\beta} (\omega-m\Omega_L, \omega - n\Omega_L).\nonumber
\end{eqnarray}
In Eq. (\ref{Floquet-Fourier}) $(m,n)$ label the Floquet modes and $(\alpha,
\beta)$ label the branch of the Keldysh contour ($\pm$) where the
respective time argument resides. Floquet modes in time-space an analogue to
Bloch modes in real space. The physical meaning of the
expansion into Floquet modes is however noteworthy, since it is the quantized absorption and emission of energy
$\hbar \Omega_L$ out of and into the classical external driving field. \\


In case of uncorrelated electrons, $U\,=\,0$, an analytical solution for the Green's
function  $G_{mn}(k,\omega)$ is found by solving the
Hamiltonian. The retarded component of $G_{mn}(k,\omega)$ is

\begin{eqnarray}
G_{mn}^{R}(k,\omega) \label{Green}
=
\sum_{\rho}
\frac
{
J_{\rho-m}\left(A_0\tilde{\epsilon}_k \right)
J_{\rho-n}\left(A_0\tilde{\epsilon}_k \right)
}
{
\omega -\rho\Omega_L - \epsilon_k + i 0^+
}.
\end{eqnarray}

$\tilde{\epsilon}_k$ represents the externally induced dispersion which is to be distinguished from the lattice dispersion
$\epsilon_k$. $J_n$ are the cylindrical Bessel functions of integer order, $A_0
= \vec{d}\cdot\vec{E}_0 $ and $\Omega_L$ characterize the external driving
laser. The physical Green's function for the optically excited band electron (LB) is derived 
\begin{eqnarray}
\label{EqGsum}
G^{R}_{\rm LB}(k,\omega)  
=
\sum_{m,n}
G_{mn}^{R}(k,\omega).
\end{eqnarray}

\subsection{Dynamical Mean-Field Theory for Electromagnetically Driven
  Semiconductors in d=3 Dimensions}

The Hamiltonian for the correlated system, $U\,\neq\,0$,
Eq. (\ref{Hamilton_we}), is solved numerically with a single-site Dynamical Mean Field Theory (DMFT) \cite{ANN,NJP,APLB} and an iterated
perturbation theory solver (IPT), see Fig. \ref{setup}(f). The
laser-band-electron Green's function $G^{R}_{\rm Lb}(k,\omega)$,
Eq. (\ref{EqGsum}), is characterized by the wave vector $k$, describing the periodicity of the lattice, the electronic energy $\omega$ and
the external driving frequency $\Omega$ as contained in the Floquet indices
$(m,n)$.  The DMFT self-consistency relation assumes the form of a $2 \times
2$  dimensional matrix equation in regular Keldysh space and in becomes  $n
\times n$ dimensional in Floquet space. The IPT is generalized to
Keldysh-Floquet form as well. The resulting numerical algorithm proves to be
efficient and stable for all values of $U$.\\ The coupling $\hat d
\cdot \vec E_0\cos(\Omega_L\tau)$ under the assumption of the Coulomb gauge
$\vec E(\tau)\,= - \frac{\partial}{\partial \tau} \vec A(\tau)$, that is in Fourier space
$\vec E(\Omega_L)\,=\,i\Omega_L\cdot\,\vec A(\Omega_L)$, generates the factor
$\Omega_L$ that cancels the $1/\Omega_L$ term in the renormalized cylindrical
Bessel function, for details see Eq. (7) of ref. \cite{NJP}. By checking the
Floquet sum we find that considering the first ten ($n=10$) Floquet modes is
sufficient within the numerical accuracy of the DMFT
\cite{LubatschEPJ2019,Symmetry}. A cut-off after a smaller number would lead to a drift
in the total energy of the system and thus it would hurt conservation laws. \\
While the quasienergy spectra of excitons and the Franz-Keldysh effect
in the non-equilibrium have been broadly investigated in
d=2 dimensions for semiconductors in the THz regime, especially for GaAs
\cite{Jauho1,Jauho2,Jauho3}, only recently the first numerical studies of
the excitonic quasienergy spectra under topological excitations in the optical
regime have been performed for ZnO with DMFT in d=3
dimensions \cite{LubatschEPJ2019,Symmetry}.  Whereas
DMFT has been so far preferentially used for typical
Mott insulators \cite{SavrasovKotliar} like NiO \cite{Fiebig,Castell}, this
DMFT study \cite{LubatschEPJ2019,Symmetry} is remarkable,
since ZnO as a wide-gapped transition metal oxide from the group IIb in the
table of the elements which may be considered to behave as a Mott insulator at
zero temperature \cite{Lu,Lu2}, especially in driven nano-structures this is
the case. ZnO is
broadly investigated for random lasing in disordered ensembles of nano-pillars and nano-grains,
i.e. Mie spheres, and a restriction to two-dimensional electronic processes
is no necessary condition in these ensembles \cite{Cao}. Two-dimensionality is
{\em a priori} also not necessary for the considering the formation of polaritons \cite{YaNature}.  
\vspace{1.5cm}
\begin{figure}[t!]
\centering
{\rotatebox{0}{\includegraphics[width=1.0\linewidth]{{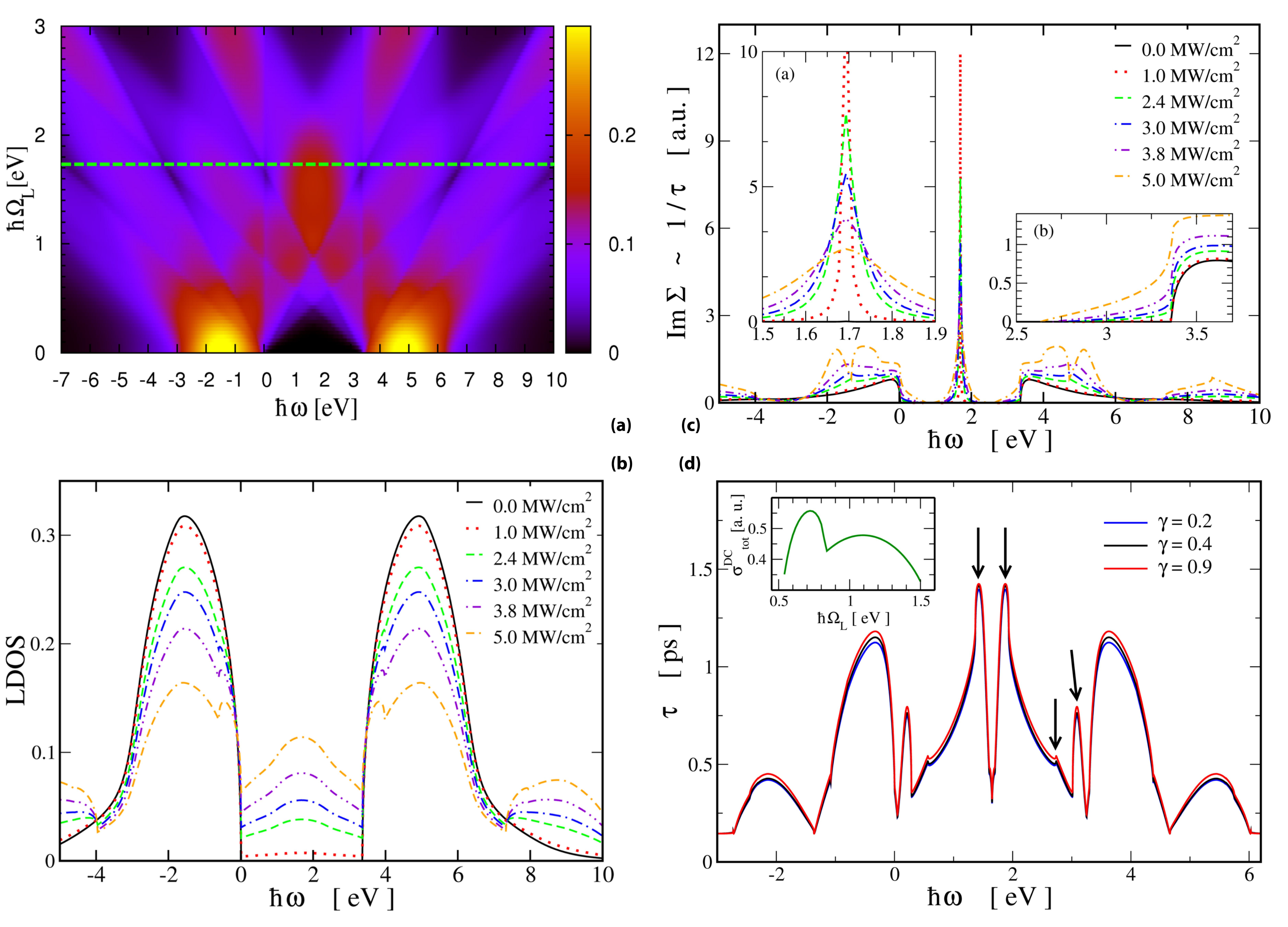}}}}
\vspace*{0.0cm}
\caption{{\bf (a)}  LDOS of bulk ZnO for varying
excitation laser frequency $\Omega_L$. The electronic gap in equilibrium is $3.38 eV$, the microscopic dipole moment equals
$|d|=2.30826288 \times 10^{-28} Cm$,  the lattice constant is $a=0,325\,
nm$. The excitation strength is $10.0 MW/cm^2$. A multitude of Floquet bands
 and sub-gaps emerge due to
the AC Stark effect. {\bf(b)} Excitonic LDOS for the driving frequency if $1.75\,eV$,
with increasing intensity, $0$ to $5.0 MW/cm^2$. Spectral weight is shifted
into the Floquet bands which cross in the gap region. Mid-gap states emerge. {\bf(c)} Inverse excitonic lifetime ${\rm Im} \Sigma$, $1/\tau$, for
parameters as in (b). (Inset a) The peak centered at $1.69 eV$ indicates small lifetimes and a
fast decay of mid-gap states (red line). (Inset b) The excitonic lifetime
near the band edges is reduced but finite. With increasing pump intensity the
near band-edge lifetime decreases but in comparison to the value in the band
it is increased.{\bf(d)} Lifetimes of exciton-polaritons in ZnO nano-pillars. 
Parameters are: pump wavelength $\lambda\,=\,710\,nm$ (1.75 eV),
cavity resonance at $\lambda\,=\,505.34\,nm$, (2.45 eV, passive
refractive index $n = 1.97182$) pump intensity $2.4 MW/cm^{2}$. Sharp
features at $0..0.5\,eV$ and $2.5..3.3\,eV$,
($\lambda\,=\,495.94..375.71\,nm$) show gain narrowing. The increase
of the coupling strength $\gamma$ of bulk matter and cavity yields rather
small effects. (Inset) The optical conductivity $\sigma_{tot}$ for
exciton-polaritons, $\gamma=0.2$, in the THz regime in the non-equilibrium is finite.}
\label{RES}
\end{figure}

\section{Results}
For driven ZnO bulk we display in Fig.\ref{RES}(a)
the local density of excitonic states (LDOS) for $10.0 MW/cm^2$ pump power and
the varying external laser frequency $\Omega_L$. The systems parameters are found
in the caption of Fig. {\ref{RES}}. A multitude of Floquet bands arises due
to the AC Stark effect and sub-gaps are formed whereas the original gap
closes. These results are computed for a cubic lattice structure and their
physical interpretation besides the formation of exciton-quasiparticles is the
generation of bands of even higher harmonics due to the symmetry of the
system. For the fixed value of $\hbar\Omega_L\,=\,1.75\,eV$, Fig.\ref{RES}(b), we
  find an increasing LDOS for excitons in the ZnO gap region. The solid black line indicates the
undriven bandstructure with a half band width of $1.7\,eV$. In the valence and
the conduction band sub gaps
towards the near-gap band edges at $-0.67\,eV$ and $4.05\,eV$ are formed,  while
at the inner band edges a step-like decrease is found. A smooth mid-gap
feature is building up with increasing pump strength. In the results for the
inverse lifetime, $1/\tau$, of excitons ${\rm Im}\Sigma$, Fig.\ref{RES}(c), which characterizes the
dynamics of the excitons we find a sharp peak at the Fermi edge, i.e. small
lifetimes of excitons as we expected it. ${\rm Im}\Sigma$ is increasing in the
gap near the equilibrium band edges. We find an increasing
LDOS for excitons with the increase of the external excitation strength. The
lifetimes of the near-edge states are therefore progressively decreased. A
plateauing of exciton lifetimes is observed in the band gap otherwise,
whereas deep in the band at e.g. $4.8\,eV$ we find for increasing external pump
intensity a crossover of the corresponding excitonic lifetimes. Here
$1/\tau$ exhibits a local maximum around a pump intensity of
$|E_0|^2\,=\,3.8\,MW/cm^2$ which then transforms with further increasing pump strength into a local minimum (orange dash-dotted line) and eventually reaches
lifetimes almost as high as in the un-driven equilibrium system (black solid
line). For the cavity coupled system, Fig. \ref{RES}(d), we find a qualitatively
different behavior. It should be noted first that the presence of the cavity
also for the weakly coupled case in the non-equilibrium always has an
influence on the full spectrum of the LDOS. This is necessarily the case for
energy and particle conservation. The semiconductor pillars with a full
diameter of $d=260.0\,nm$ are chosen with a
geometrical Mie resonance as a single whispering gallery mode on the inner
jacket of the resonator. The coat thickness may vary with the fineness of the ZnO
material. As such it may determine the Q-factor of the single Mie resonator in
principle \cite{Koch}, however in the non-equilibrium regime and in the corresponding
physical time range microscopic interactions in the sense of impurity scattering in the
selvedge have no fundamental influence. The Mie resonance of $\hbar\omega_0 \!\!= 2.45\,eV$ corresponds to the
wavelength $\lambda\,=\,505.34\,nm$ \cite{FrankPhysRevResearch}, effectively
this whispering gallery is populated in the resonance by photons. Thus the
classical picture of the Mie resonance \cite{Mie} of a sole geometric
resonance of otherwise passive matter is replaced here by the Fano resonance
\cite{Fano} of the single mode with the continuum of excitons which is
embedded in a Floquet matrix \cite{Floquet,Grifoni,PRB,FrankANN}. The Floquet matrix, see
Fig. \ref{setup}(a), represents mathematically the physics of high amplitude
excitations and higher harmonics \cite{Faisal} under the presumption of energy conservation. 
We display the exciton-polariton lifetimes for 
$(\!g/t)\!\!=\!\! 0.2, 0.4, 0.9$ at zero temperature and at half
filling. The coupling strength as such is characteristic for the
formation of polaritons \cite{Lamata}. The results are derived for an
external optical pump intensity of $2.4 MW/cm^2$ which on the other side is a typical pump strength for solid state
random lasers \cite{NJP14,SREP15,ApplSci,KaltPSS}.  We find a sharp dip in the lifetimes of
polaritonic  states
at the Fermi energy.  In agreement with the results for the bulk we also find
the characteristic features around the Fermi edge. The dip of $1/\tau$ at
$4.8\,eV$ for the driving intensity of $2.4 MW/cm^2$ for bulk persists in the
full cavity coupled system, Fig. \ref{RES}(d). The magnitude of the exciton-polariton lifetimes in the
marked features of gap states is $1.45$ ps (for $1.4\,eV$ and $1.9\,eV$), $0.6$ ps (for $2.75\,eV$), and $0.75$
ps (for $3.1\,eV$). The peak of the exciton-polariton lifetimes in the original
band at $3.65\,eV$ is derived as $1.25$ ps. The  peaks in the results
for the polaritonic lifetime (black arrows in Fig. \ref{RES}(d)) are all
emerging quite far off from the cavity's resonance at $\lambda\,= \,2.45\,eV$
which is a proof of the dynamic Fano coupling in the driven non-equilibrium
system. The peaks can be interpreted as a splitting into
several polariton branches which for the given parameters of random lasers may
define the laser spectrum of a random polariton laser consisting of ZnO Mie
resonators. The result for the optical DC-conductivity  $\sigma_{tot}$ as the
materials characteristics
in the non-equilibrium is displayed for $\gamma\,=\,0.2$ in the
inset of Fig. \ref{RES}(d) in the THz region and the same parameters otherwise.
$\sigma_{tot}$ assumes a finite value. The
pronounced minima in $\sigma_{tot}$ are another signature of gain and gain
narrowing in the full polaritonic spectrum. Our results of lifetimes of
excitons and exciton-polaritons are corroborated by
the experimental findings of laser spectra in several setups of ZnO Mie
resonators as well as in ZnO random lasers \cite{NJP14,SREP15,ApplSci,
    KaltPSS,SUN,DUAN,srep20581} where in either experiment stable lasing modes at 3.13 eV have been
  found. In either system the spectral result is attributed to the coupling of the
  excited electronic subsystem of ZnO bulk to the resonator mode of the
  single scatterer. Refs. \cite{SUN,DUAN,srep20581} interpret their findings
  as exciton-polaritons.

\section{Conclusions}

We presented in this article results for the local density of states and the lifetimes of
excitons and of exciton-polaritons in excited ZnO Mie resonators in the
non-equilibrium. The sharp spectral features of the polaritonic lifetimes
are in clear qualitative and quantitative contrast to the results for the exciton lifetimes.
Furthermore they are in qualitative and in quantitative agreement with
experimental results of ZnO polariton lasers on the one hand and on the other
hand they also agree with
results for the spectrum and the life times of solid state random lasers. We
conclude that a random polariton laser can be built as an application of ZnO
Mie resonators in the non-equilibrium.

\vspace{6pt} 



\authorcontributions{All authors contributed equally to this work. All authors
wrote and reviewed the manuscript.}

\acknowledgments{The authors thank H. Cao, P. Guyot-Sionnest, H. Kalt,
  A.-P. Jauho, M. Richard, J.K. Freericks and
G. Kotliar for fruitful discussions.}


\reftitle{References}



\begin{thebibliography}{999}
\bibitem{YaNature} Yamamoto, Y.; Half-matter, half-light amplifier, {\em
    Nature} {\bf 2000}, 405, 629-630.

\bibitem{Imamoglu} Imamoglu, A.; Ram, R. J.; Pau, S.; Yamamoto, Y.; Nonequilibrium
  condensates and lasers without inversion: Exciton-polariton lasers, 
  {\em Phys. Rev. A} {\bf 1996}, 53, 4250-4253.

\bibitem{Kaliteevski} Zamfirescu, M.; Kavokin, A.; Gil, B.; Malpuech, G.; 
  Kaliteevski, M.; ZnO as a material mostly adapted for the realization of
  room-temperature polariton lasers, {\em Phys. Rev. B} {\bf 2002}, 65, 161205(R).

\bibitem{YaScience} Deng, H.; Weihs, G.; Santori, C.; Bloch, J.; Yamamoto, Y.; 
  Condensation of semiconductor microcavity exciton polaritons, 
  {\em Science} {\bf 2002}, 298, 199-202.

\bibitem{Dang} Kasprzak, J.; Richard, M.; Kundermann, S.; Baas, A.; Jeambrun,
  P.; 
  Keeling, J. M.; Marchetti, F. M.; Szymanska, M. H.; Andre, R.; Staehli,
  J. L.; 
  Savona, V.; Littlewood, P. B.; Deveaud, B.; Dang, L. S.; Bose-Einstein
  condensation of exciton polaritons, {\em Nature} {\bf 2006}, 443, 409-414.

\bibitem{Hoefling} Schneider, C.; Rahimi-Iman, A.; Kim, Y. N.; Fischer, J.; 
  Savenko, I. G.; Amthor, M.; Lermer, M.; Wolf, A.; Worschech, L.; 
  Kulakovskii, V. D.; Shelykh, I. A.; Kamp, M.;  Reitzenstein, S.; Forschel,
  A.; 
  Yamamoto, Y.; H\"ofling, S.; An electrically pumped polariton laser, {\em
    Nature} {\bf 2013}, 497, 348-352.

\bibitem{Li}  F. Li, L. Orosz, O. Kamoun, S. Bouchoule, Ch. Brimont,
  P. Disseix, T. Guillet, X. Lafosse, M. Leroux, J. Leymarie, G. Malpuech,
  M. Mexis, M. Mihailovic, G. Patriarche, F. Reveret, D. Solnyshkov,
  J. Zuniga-Perez, Fabrication and characterization of a room-temperature ZnO
  polariton laser, {\em Appl. Phys. Lett.} {\bf 102}, 191118 (2013).

\bibitem{Guillet}
Guillet, T.; Mexis, M.; Levrat, J.;  Rossbach, G.; Brimont, Ch.; Bretagnon,
T.; Gil, B.; Butte, R.; Grandjean, N.; Orosz, L.; Reveret, F.; Leymarie, J.; 
Zuniga-Perez, J.; Leroux, M.; Semond, F.; Bouchoule, S.; Polariton lasing in a
hybrid bulk ZnO microcavity, {\em Appl. Phys. Lett.} {\bf 2011}, 99, 161104.

\bibitem{Bajoni} Bajoni, D.; Polariton lasers. Hybrid light-matter lasers
  without inversion, {\em J. Phys. D: Appl. Phys.} {\bf 2012}, 45, 313001. 

\bibitem{Cao} Cao, H.; Zhao, Y. G.; Ho, S. T.; Seelig, E. W.;  Wang, Q. H.;
  Chang, R. P. H.; Random Laser Action in Semiconductor Powder, {\em
    Phys. Rev. Lett.} {\bf 1999}, 82, 11 2278.

\bibitem{Sasaki} Niyuki, R.; Fujiwara, H.; Nakamura, T.; Ishikawa, Y.; 
  Koshizaki, N.; Tsuji, T.; Sasaki, K.; Double threshold behavior in a
  resonance-controlled ZnO random laser, {\em APL Photonics} {\bf 2017}, 2, 036101.

\bibitem{Deng1} Lu, T.-C.; Lai, Y.-Y.; Lan, Y.-P.; Huang, S.-W.; Chen, J.-R.; 
Wu, Y.-C.; Hsieh, W.-F.; Deng, H.; Room temperature polariton lasing vs. photon
lasing in a ZnO-based hybrid microcavity, {\em Opt. Express} {\bf 2012}, 20, 5530.

\bibitem{Deng2} Kim, S.; Zhang, B.; Wang, Z.; Fischer, J.; Brodbeck, S.; Kamp,
  M.; Schneider, C.; H\"ofling, S.; Deng, H.; Coherent Polariton Laser, {\em
    Phys. Rev. X} {\bf 2016}, 6, 011026.

\bibitem{LubatschEPJ2019} Lubatsch, A.; Frank, R.; Evolution of Floquet
  Topological Quantum States in Driven Semiconductors, {\em Eur. Phys. J. B}
  {\bf 2019}, 92: 215, 92 (9), doi:10.1140/epjb/e2019-100087-0.


\bibitem{Symmetry} Lubatsch, A.; Frank, R.; Behavior of Floquet Topological
  Quantum States in Optically Driven Semiconductors. {\em Symmetry} {\bf
    2019}, 11, 1246; doi: 10.3390/sym11101246.

\bibitem{Jauho1} Jauho, A.-P.; Johnsen, K.; Dynamical Franz-Keldysh Effect, {\em
    Phys. Rev. Lett.} {\bf 1996}, 76, 24, 4576-4579.

\bibitem{Jauho2} Nordstrom, K.B.; Johnsen, K.; Allen, S.J.; Jauho, A.-P.;
  Birnir, B.; Kono, J.; Noda, T.; Akiyama, H.; Sakaki, H.;  Excitonic
  Dynamical Franz-Keldysh Effect, {\em
    Phys. Rev. Lett.} {\bf 1998}, 81, 2, 457-460.


\bibitem{Jauho3} Johnsen, K.; Jauho, A.-P.; Quasienergy Spectroscopy of
  Excitons, {\em
    Phys. Rev. Lett.} {\bf 1999}, 83, 6, 1207-1210.


\bibitem{SavrasovKotliar}Savrasov, S. Y.; Kotliar, G.; Linear Response
  Calculations of Lattice Dynamics in Strongly Correlated Systems,  {\em
    Phys. Rev. Lett.} {\bf 2003}, 90 (5) 056401-1.

  
\bibitem{Fiebig} Fiebig, M.; Fr\"ohlich, D.; Lottermoser, Th.; Pavlov, V. V.;
  Pisarev, R. V.; Weber, H.-J.; , Second Harmonic Generation in the
  Centrosymmetric Antiferromagnet NiO, {\em Phys. Rev. Lett.} {\bf 2001}, 87 (13), 137202-1.
  
\bibitem{Castell} Castell, M. R.; Wincott, P. L.; Condon, N. G.; Muggelberg,
  C.; Thornton, G.; Dudarev, S. L.; Sutton, A. P.; Briggs, G. A. D.;
  Atomic-resolution STM of a system with strongly correlated
  electrons:NiO(001) surface structure and defect sites, {\em Phys. Rev. B} {\bf
    1997}, 55 (12), 7859.

  \bibitem{Koch} Gibbs, H. M.; Khitrova, G.; Koch, S. W.; Exciton–polariton
  light–semiconductor coupling effects. {\em Nat. Phot.} {\bf 2011}, 5, 275-282.



\bibitem{Lu} Chang, P.-C.; Lu, J. G.; Temperature dependent conduction and UV
  induced metal-to-insulator transition in ZnO nanowires, {\em
    Appl. Phys. Lett.} {\bf 2008}, 92, 212113.

\bibitem{Petrosyan} Aghamalyan, N. R.; Aslanyan, T. A.; Vardanyan, E. S.; 
   Kafadaryan, Y. A.; Hovsepyan, R. K.; Petrosyan, S. I.; Poghosyan, A. R.; 
  Metal-insulator electronic phase transitions in wide-gap ZnO
  semiconductors.  {\em
    Journal of Contemporary Physics} (Armenian Academy of Sciences), {\bf
    2012}, 47, 6, 275-281.

\bibitem{Lu2} Chang, P.-C.; Chien, C.-J.; Stichtenoth, D.; Ronning, C.; Lu,
  J. G.; 
  Finite size effect in ZnO nanowires, {\em
    Appl. Phys. Lett.} {\bf 2007}, 90, 113101.

\bibitem{Shimada}  Shimada, R.; Xie, J.; Avrutin, V.; \"Ozg\"ur, \"U.; 
  Morkovic, H.; Cavity polaritons in ZnO-based hybrid microcavities, {\em
    Appl. Phys. Lett.} {\bf 2008}, 92, 011127. 

\bibitem{SUN} Dai, J.; Xu, C. X.; Sun, X. W.; Zhang, X. H.; Exciton-polariton
  microphotoluminescence and lasing from ZnO whispering-gallery mode
  microcavities, {\em
    Appl. Phys. Lett.} {\bf 2011}, 98, 161110.

\bibitem{DUAN} Duan, Q.; Xu, D.; Liu, W.; Lu, J.; Zhang, L.; Wang, J.; Wang,
  Y.; Gu, J.; Hu, T.; Xie, W.; Shen, X.; Chen, Z.; Polariton lasing of quasi-whispering
  gallery modes in a ZnO microwire, {\em
    Appl. Phys. Lett.} {\bf 2012}, 103, 022103.

\bibitem{JMATCHEMA} Razavi-Khosroshahi, H.; Edalati, K.; Wu, J.; Nakashima,
  Y.; Arita, M.; Ikoma, Y.; Sadakiyo, M.; Inagaki, Y.; Staykov, A.; Yamauchi,
  M.; Horita, Z.;  Fuji, M.; High-pressure zinc oxide phase as visible-light-active
  photocatalyst with narrow band gap, {\em J. Mater. Chem. A}, {\bf 2017}, 5, 20298-20303.

 \bibitem{ZnOMono1} Huang, F.; Lin, Z.; Lin, W.; Zhang, J.; Ding, K.; Wang,
   Y.; Zheng, Q.; Zhan, Z.; Yan, F.; Chen, D.;  Lv, P.; Wang, X.; 
 Research progress in ZnO singlecrystal: Growth, scientific understanding, and
 device applications, {\em Chinese
    Science Bulletin} {\bf 2014}, 59, 1235.

\bibitem{ZnOMono2} Park, W. I.; Jun, Y. H.; Jung, S. W.; Yi, G.-C.; Excitonic
  emissions observed in ZnO single crystal nanorods, {\em
    Appl. Phys. Lett.} {\bf 2003}, 82, 6 964-966.

\bibitem{Rocksalt1} Fritsch, D.; Schmidt, H.; Grundmann, M.; Pseudopotential band
  structures of rocksalt MgO, ZnO, and Mg1-xZnxO, {\em
    Appl. Phys. Lett.} {\bf 2006}, 88, 134104.

\bibitem{Rocksalt2} Dixit, H.; Saniz, R.; Lamoen, D.; Partoens, B.; The
  quasiparticle band structure of zincblende and rocksalt ZnO, {\em
    J. Phys. Condens. Matter} {\bf 2010},  22, 125505.

\bibitem{vanBlaaderen} Koster, R. S.; Changming, M. F.; Dijkstra, M.; van
  Blaaderen, A.; van Huis, M. A.; Stabilization of Rock Salt ZnO Nanocrystals
  by Low-Energy Surfaces and Mg Additions: A First-Principles Study, 
{\em J. Phys. Chem. C} {\bf 2015}, 119, 5648-5656 doi: 10.1021/jp511503b.


\bibitem{NJP14}  Lubatsch, A.; Frank, R.; Coherent transport and symmetry
  breaking - laser dynamics of constrained granular matter, {\em New J. Phys.}
  {\bf 2014}, 16, 083043.


\bibitem{SREP15} Lubatsch, A.; Frank, R.; Tuning the Quantum Efficiency of Random
  Lasers - Intrinsic Stokes-Shift and Gain, {\em Scientific Reports} {\bf 2015}, 5(1)
    17000. 

\bibitem{ApplSci} Lubatsch, A.; Frank, R.; A Self-Consistent Quantum Field
    Theory for Random Lasing, {\em Appl. Sci.} {\bf 2019}, 9, 2477
    doi:10.3390/app9122477.

\bibitem{Mie} Mie, G. Beitr\"age zur Optik tr\"uber Medien, speziell kolloidaler
  Metall\"osungen, {\em Ann Phys.} (Berlin) {\bf 1908}, 4, 25, 377-445.

\bibitem{Fano} Fano, U.; Effects of Configuration Interaction on Intensities
  and Phase Shifts, {\em Phys. Rev.} {\bf A} {\bf 1961}, 124, 1866.

\bibitem{Floquet} Floquet, G. Sur les équations différentielles linéaires à coefficients périodiques.  {\em Ann. l' Ecole Norm. Sup.} {\bf
    1883}, \emph{12}, 47-88.  


  \bibitem{Grifoni} Grifoni, M.;  H\"anggi, P. Driven quantum tunneling. {\em Phys. Rep.} {\bf
    1998}, \emph{304}, 229--354.     

\bibitem{PRB} Frank, R. Coherent control of Floquet-mode dressed plasmon polaritons. {\em Phys. Rev. B} {\bf 2012}, \emph{85}, 195463.                  

\bibitem{FrankANN} Frank, R. Non-equilibrium polaritonics - Nonlinear effects and optical switching. {\em Ann. Phys.}  {\bf 2013}, \emph{525},          
66--73.

\bibitem{Kotliar} Georges, A.; Kotliar, G.; Krauth, W.; Rozenberg, M.J.                 
  Dynamical mean-field theory of strongly correlated fermion systems and the limit of infinite dimensions. {\em Rev. Mod. Phys.} {\bf 1996}, \emph{68}, 13.


\bibitem{ANN} Lubatsch, A.; Kroha, J. Optically driven Mott-Hubbard systems           
  out of thermodynamical equilibrium. {\em Ann.~Phys.}  {\bf
    2009}, \emph{18}, 863--867.

\bibitem{NJP} Frank, R. Quantum criticality and population trapping of
  fermions by non-equilibrium lattice modulations. {\em New J. Phys.} {\bf
    2013}, \emph{15}, 123030.

\bibitem{APLB} Frank, R. Population trapping and inversion in ultracold Fermi
  gases by excitation of the optical lattice-Non-equilibrium Floquet - Keldysh description. {\em Appl. Phys. B} {\bf 2013}, \emph{113}, 41--47.                  


   \bibitem{Faisal} Faisal, F.H.M.; Kaminski, J.Z. Floquet-Bloch theory of
     high-harmonic generation in periodic structures. {\em \mbox{Phys. Rev.
         A}} {\bf 1997}, \emph{56},  748.

       \bibitem{FrankPhysRevResearch} Lubatsch, A.; Frank, R.;  Self-consistent quantum field theory for the characterization of complex random
media by short laser pulses, {\em \mbox{Phys. Rev. Research}} {\bf 2020},
\emph{https://arxiv.org/abs/2001.02742 accepted, in press}.


 \bibitem{Lamata} Forn-Diaz, P.; Lamata, L.; Rico, E.; Kono, J.; Solano, E.;
  Ultrastrong coupling regimes of light-matter interaction. {\em
    Rev. Mod. Phys.}  {\bf 2019}, \emph{91}, 025005-1. 


\bibitem{Zhang}  Zhang, F. C.; Rice, T. M.; Effective Hamiltonian for the
  superconducting Cu oxides, {\em Phys. Rev. B} {\bf 1998},
   37, 7, 3759.

\bibitem{Jarell} Jarell, M.;  Freericks, J.K.; Pruschke, Th.; Optical conductivity
  of the infinite-dimensional Hubbard model, {\em Phys. Rev. B}
  {\bf 1995}, 51, 17, 11704.


\bibitem{Eder} Eder, R.; van den Brink, J.; Sawatzky, G. A.; Intersite Coulomb
  interaction and Heisenberg exchange, {\em Phys. Rev. B}
  {\bf 1996}, 54, 2, R 732(R).

\bibitem{KaltPSS} Kalt, H.; Fallert, J.; Dietz, R. J. B.; Sartor, J.; 
  Schneider, D.; Klingshirn, C.; Random lasing in nanocrystalline ZnO powders,
  {\em Phys. Status Solidi B} {\bf 2010}, 247,
  No. 6, 1448-1452.

\bibitem{srep20581} Lai, Y.-Y.; Chou, Y.-H.; Lan, Y.-P.; Lu, T.-C.; Wang, S.-C.;
  Yamamoto, Y.; Crossover from polariton lasing to exciton lasing in a strongly
  couples ZnO microcavity {\em  Scientific Reports}, {\bf 2015} 6, 20581.
\end{thebibliography}



\end{document}